\documentclass{iau} 
\usepackage{graphicx}

\title[JD 11.~~IMF: Progress and Challenges] 
{An Observational Perspective of the IMF: Progress and Challenges}

\author[Stella S. R. Offner]   
{Stella S. R. Offner$^1$}

\affiliation{$^1$University of Massachusetts, \\
Amherst, MA, USA \\ email: {\tt soffner@astro.umass.edu}}

\pubyear{2015}
\volume{315}  
\jname{From Interstellar Clouds to Star-Forming Galaxies: Universal Processes?}
\editors{P. Jablonka, F. Van der Tak, \& P. Andr/'e, eds.}
\begin{document}

\maketitle

\begin{abstract}
The stellar initial mass function (IMF) is a fundamental astrophysical quantity that impacts a wide range of astrophysical problems from heavy element distribution to galactic evolution to planetary system formation. However, the origin and universality of the IMF are hotly debated both observationally and theoretically. I review recent observations of the IMF across a variety of environments. These suggest the IMF is surprisingly invariant between star-forming regions, star clusters, and spiral galaxies but that it may also vary under extreme conditions, including within the Galactic center and early type galaxies. 
\end{abstract}

\firstsection 

\section{Introduction}

The shape of the stellar initial mass function (IMF) has been debated for more than fifty years. Since the underlying physics of star formation is complex and nonlinear, however, our view of the IMF rests on observational studies of individual stars (Bastian et al. 2010). Observational biases, dynamical evolution and poor statistics complicate the situation, and the result is an IMF constructed from a jigsaw of old and young stellar populations, dense and open star clusters, and photometric and spectroscopic approaches. Of fundamental importance is whether the IMF is universal (Offner et al. 2014). This problem has reached a new urgency with the proliferation of studies of massive elliptical galaxies, with properties very unlike our own, in which individual stars cannot be resolved.  Here I briefly review observational studies that form the basis of our knowledge of the IMF and discuss recent progress and current challenges in this field.

\begin{figure}[b]
\begin{center}
 \includegraphics[width=3.5in]{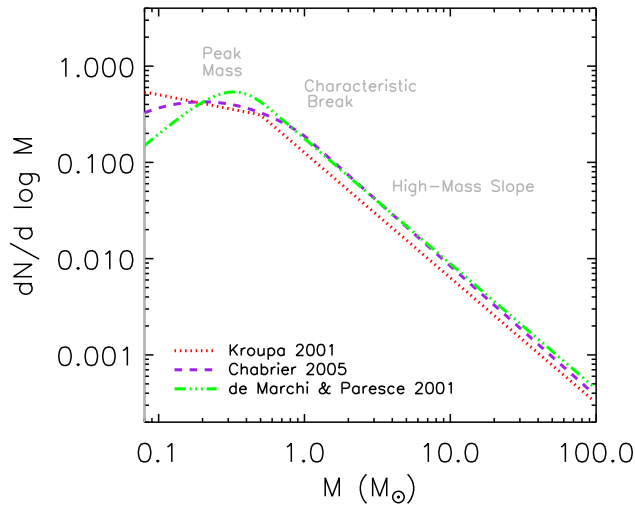}
 \vspace*{-0.2 cm}
 \caption{Three proposed functional forms for the IMF, which are calibrated using studies of the population of field stars and star clusters. }
   \label{fig1}
\end{center}
\end{figure}

\section{Functional Forms for the IMF}\label{forms}

The IMF is typically expressed as a number of stars per linear or logarithmic bin in mass. The functional form for the IMF is derived empirically from observational studies of Milky Way field stars or star clusters. Observational uncertainties are sufficiently large (see \S\ref{resolpop}) that a variety of analytical forms put forth in the literature are statistically consistent with the data. An early determination of the IMF by Salpeter (1955) found that the number of stars $dN/ d {\rm log} M \propto M^{-1.35}$.  Subsequent studies identified a peak mass around a few tenths of a solar mass. This low-mass turnover can be expressed by broken power-laws (Kroupa 2001), a log-normal  (Chabrier 2003, 2005), or an exponential (de Marchi \& Paresce 2001).  Table \ref{tab1} summarizes these common functional forms.  As shown in Figure \ref{fig1}, sixty years later the Salpeter IMF is widely adopted for stars with masses above 1 $M_\odot$. Meanwhile, the dispersion in the placement of the peak underscores the current level of uncertainty at lower stellar masses. The curves diverge below the brown dwarf mass limit, 0.08 $M_\odot$, which highlights the magnitude of the disagreement regarding very low mass objects. The IMF is sometimes given as a system IMF, which ignores the individual stellar components of binary and higher order star systems in favor of counting the total system mass. However, the single-star and system IMF shapes appear similar (Chabrier 2005).
 
\begin{table}
  \begin{center}
  \caption{Common functional forms for the IMF.}
  \label{tab1}
 {\scriptsize
  \begin{tabular}{|l|c|c|c|}\hline 
{\bf Citation} & {\bf Peak}  & {\bf High-Mass} & {\bf $dN/ d {\rm log} M$}$^1$ \\ 
& {\bf Mass($M_{\odot}$)} & {\bf Slope, $\alpha$} &  \\ \hline
Salpeter 1955 		&  ...	& -1.35 &  $ M^{-1.35}$\\ 
Kroupa 2001 			&  0.08	& -1.3 & $0.01<M<0.08$: $M^{0.7}$, $0.08 \leq M<0.5$: $M^{-0.3}$,  \\ 
			                  &  	         &    &   $M \geq 0.5$: $M^{-1.3}$  \\ 
de Marchi \& Paresce 2001 & 0.3 	& -1.3 & $M^{-1.3} (1-{\rm exp}[-(M/0.3)^{2.6}])$ \\
Chabrier 2003 		&  0.08	&  ...$^2$ &  $M \leq 1.0: {\rm exp}(-[{\rm log}_{10}(M)-{\rm log}_{10}(0.08)]^2/(2\times0.69^2))$   \\ 
Chabrier 2005 		&  0.2	& -1.35 &  $M \leq 1.0: {\rm exp}(-[{\rm log}_{10}(M)-{\rm log}_{10}(0.2)]^2/(2\times0.55^2))$,   \\ 
		&  	&  &    $M \leq 1.0:$ $M^{-1.35}$ \\ 
\hline
  \end{tabular}
  } 
 \end{center}
\vspace{1mm}
 \scriptsize{
 {\it Notes:}\\
  $^1$ Mass $M$ in units of $M_{\odot}$. Forms are unnormalized. \\ 
  $^2$ Chabrier (2003) does not provide the IMF form above 1 $M_{\odot}$, but extragalactic studies applying this form commonly assume a Salpeter slope. \\ 
  }
\end{table}

\section{Studies of Resolved Populations}\label{resolpop}

\subsection{Observational Procedures and Uncertainties}

Studies of nearby resolved stars provide the most direct avenue for determining the IMF.  However, even when individual stars may be measured, a number of steps are involved, each of which introduces new complications and challenges.
The first step of the process entails identifying a volumetrically complete sample of stars and measuring their luminosities. Star clusters are ideal targets since they are spatially concentrated and tend to have similar ages and distances. Next, the luminosity function (LF) must be converted to a present-day mass function using models for the relationship between magnitude and stellar mass. Finally, this ``present-day" mass function must be corrected for stars missing due to dynamical or stellar evolution as well as for unresolved binary.

Despite proximity, studies of resolved population suffer from significant systematic uncertainties. These uncertainties can be divided into those associated with measuring the LF and those associated with theoretical models. Foremost, completeness and contamination can significantly bias the LF. Background galaxies and non-member stars (for a cluster) must be identified and removed. Field star LFs must be corrected for Malmquist bias and distance uncertainties. Star clusters require follow-up proper-motion studies to accurately determine cluster membership. Patchy or differential dust extinction afflicts studies of young regions ($<10$ Myr). These combined corrections significantly reduce the number of sources, and thus, the remaining sample may have poor number statistics, requiring special care. The centers of dense clusters require corrections for crowding. Finally, since half of all stars are in binary systems (Duch{\^e}ne \& Kraus 2013), statistical corrections must be applied to account for unresolved multiplicity. Stellar multiplicity is observed to vary with age in star forming regions, however, whether the multiplicity distribution varies with environment is poorly constrained.

Theoretical uncertainties are equally critical. Mass-magnitude models are sensitive to underlying assumptions about stellar evolution, and the model choice can result in order unity differences in inferred stellar properties (e.g., Da Rio et al. 2012). Pre-main sequence models for stars $<10$ Myr are especially sensitive to the initial assumptions for the starting stellar properties and how mass is deposited on the star (Hosokawa et al. 2011, Baraffe et al. 2012).  Low-mass stars are magnetically active, which can impact measured temperatures and colors (Stassun et al. 2014). Finally, corrections for dynamical interactions, which create mass segregation and can preferentially remove low-mass members, also add uncertainty to older populations. 

\begin{figure}[b]
\vspace{-0.1 cm}
\begin{center}
 \includegraphics[width=3.5in]{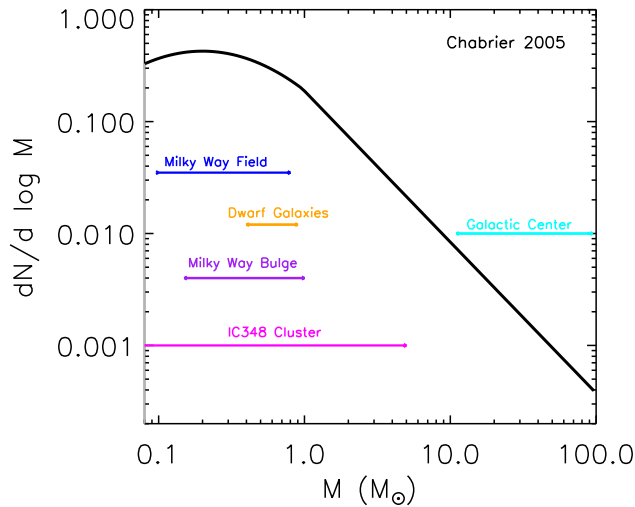}
 \vspace*{-0.2 cm}
 \caption{Chabrier IMF over-plotted with horizontal lines indicating stellar mass ranges for specific resolved population studies: the Milky Way field (Bochanski et al. 2010), local dwarf galaxies (Geha et al. 2013),  the Milky Way bulge (Calamida et al. 2015),  the Galactic center (Lu et al. 2013), the young nearby cluster IC348 (Alves de Oliveira et al. 2013), and a selection of intermediate mass star clusters in M31 (Weisz et al. 2015). }
   \label{fig2}
\end{center}
\end{figure}

\begin{figure}[b]
\vspace{-0.1 cm}
\begin{center}
\includegraphics[width=3.5in]{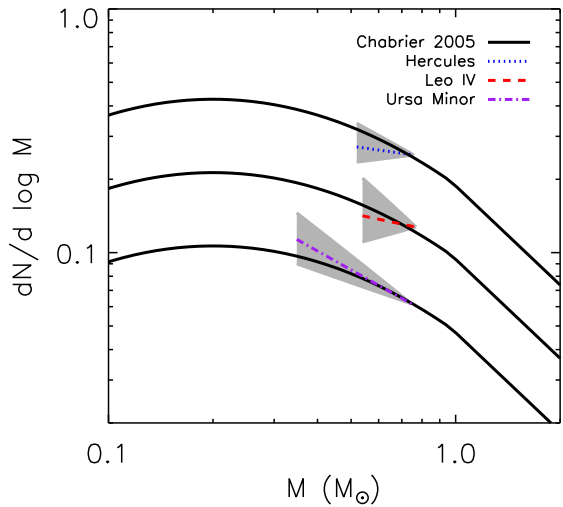}  
 \vspace*{-0.2 cm}
 \caption{ The IMF slopes obtained for three local dwarf galaxies using resolved population studies (Geha et al. 2013,  Wyse et al. 2002):  $\alpha=-0.2^{+0.4}_{-0.5}$ (Hercules), $\alpha=-0.3 \pm 0.8$ (Leo IV), and $\alpha =-0.8$ (Ursa Minor). The dwarfs have visual magnitudes $M_v=$ -6.2, -5.5 and -9.2, respectively. The $1\sigma$ uncertainty in the slope is indicated by the grey shaded area, where $15\%$ uncertainly is adopted for the latter case based on the LF error. The lines are offset for clarity.}
   \label{fig3}
\end{center}
\end{figure}

\subsection{Universality and Outliers}

Whether or not the IMF is universal hinges on the existence of {\it systematic} changes in the IMF form as a function of some underlying physical variable. If a given observed IMF is a statistical sample drawn from a universal function, then some deviations are expected, especially for small stellar populations. For example, in very rare cases a high-mass star may form in relative isolation (Offner et al. 2014, references therein). 

As shown in Figure \ref{fig2}, many studies probe only a narrow range of masses and do not fully sample the IMF.  Consequently, the size of the observational errors, as noted above, and the difficulty of deriving an IMF from a present-day mass function prohibit a clear determination of universality. However, despite this challenge there is broad consensus that the Milky Way field population, bulge population, and young clusters within the Milky Way all share a common IMF (Bochanski et al. 2010, Calamida et al. 2015, Alves de Oliveira et al. 2013). Likewise, a recent comprehensive study of young intermediate mass star clusters within M31 confirmed an invariant, universal high-mass IMF slope as a function of cluster mass, size and age (Weisz et al. 2015). The bottom line is that {\it the IMF is remarkably invariant across a broad range of environments within observational uncertainties. }


Despite a broad consensus for the IMF form among resolved populations, debate continues for a few individual regions. For example, Taurus, a young low-mass star forming region, appears to have an excess of solar mass stars (Luhman 2012). It is unclear whether this is due to small statistical sampling or has some physical basis. The Taurus region is not physically distinct compared to other young regions, which don't exhibit evidence for IMF variations (e.g., Figure 2 in Offner et al. 2014), so there's no obvious physical rational for variation. 

Within the Milky Way there is one very conspicuous extreme star-forming environment: the Galactic center. Indeed the inner parsec of the Milky Way appears to have a present-day mass function with an excess of high-mass stars (e.g., Nayakshin \& Sunyaev 2005, Bartko et al. 2010). Several independent groups have found that the high-mass slope is shallower than the Salpeter slope (e.g., $\alpha =-0.7\pm0.2$ versus $\alpha=-1.35$ for $M_* \ge 10 M_\odot$, Lu et al. 2013). However, most recently Lu et al. (2013) find that after correcting for extinction only one bin in the $Kp$ band LF deviates by more than $2\sigma$ from a Milky Way IMF.  Considering the extreme temperatures, densities and radiation field in addition to uncertainties in age, dynamics, and statical sampling, the degree of variation is surprisingly small.  

Recently, a few groups have constructed mass functions from resolved studies of ultra-faint dwarf galaxies. The smallest dwarfs are dark matter dominated and comprised of mere hundreds to thousands of detectible stars.  Given their small size, the observed stars likely formed prior to reionization, at which point their star formation was quenched. Considered together, these environments may show some systematic IMF change and appear to have relatively more high-mass stars than expected (Geha et al.~2013). However, as shown in Figure \ref{fig3}, the observations probe a narrow region near the peak of the IMF, where the slope transitions from $\alpha = -1.35$ to $\alpha = 0$. The mass ranges span as little as $0.5 M_\odot < M_* < 0.8 M_\odot$.  Wyse et al. (2002) in fact concluded that the IMF slope of the Ursa Minor dwarf spheroidal was consistent with that of globular clusters, and hence, supported an invariant IMF. Although it's possible that these data in combination with extragalactic unresolved populations signal a broader IMF trend with stellar velocity dispersion (\S\ref{unresolpop}), the significance of the effect is very tentative given the magnitude of the observational errors, which place the data within $1\sigma$ of the Milky Way IMF.  


\section{Studies of Unresolved Populations}\label{unresolpop}

\subsection{Observational Procedures and Uncertainties}

 IMF studies of unresolved stellar populations probe the large-scale mass distribution and integrated light and then attempt to back out the underlying stellar population.  Compared to resolved population studies, these approaches have less sensitivity to particular masses and, consequently, details of the functional form are often lost.  
 Consequently, extragalactic stellar populations are generally characterized as being similar to a Milky Way IMF (e.g., Kroupa 2001 or Chabrier 2003), Salpeter IMF ($\alpha=-1.35$ down to 0.1$M_\odot$), ``top-heavy" IMF ($\alpha < -1.35$ down to 0.1$M_\odot$), or ``bottom-heavy" IMF ($\alpha > -1.35$ down to 0.1$M_\odot$). The latter are often referred to as ``bottom-light" and ``top-light" IMFs, respectively.  By definition, a Salpeter IMF is also slightly bottom-heavy compared to a Milky Way IMF.
 

Stellar population synthesis (SPS) plays an important role in most unresolved population studies. Given some functional form for the stellar mass distribution, $\Phi(M)$, SPS calculates the integrated flux as a function of wavelength: $F_{\lambda}= \int \Phi(M)f_{\lambda} dM$, where $f_\lambda (M)$ is the spectrum of a star with mass $M$. This approach requires assumptions for the age and metallicity of the population. Together with mass determinations from gravitational lensing (Shu et al. 2015) or stellar dynamics (Cappellari et al. 2012), it is possible to construct the mass-to-light ratio. 
High mass-to-light ratios indicate a larger fraction of low-mass stars and correspond to a steeper, bottom-heavy IMF, while lower mass-to-light ratios indicate a relative excess of high-mass stars and correspond to a top-heavy IMF. 

One caveat to integrated approaches is that the total mass is the sum of the stellar and dark matter masses. Extracting the stellar mass requires modeling the underlying distribution of dark matter, which cannot be directly measured. In principle, any variations in the IMF can be removed with an appropriate choice of dark matter profile. However, this is generally discouraged since nontrivial revisions to cosmological models would be necessary in some cases (e.g., Cappellari et al. 2012).


An alternative form of SPS involves modeling gravity sensitive spectral features found in stellar atmospheres (Conroy \& van Dokkum 2012). This technique is promising since it provides a means of assessing the relative numbers of dwarf and giant stars that is independent of the total mass distribution. In addition,  it reproduces the Milky Way IMF found in globular clusters (van Dokkum \& Conroy 2011). However, spectral SPS depends on a large number of parameters, including initial elemental abundances and the star formation history. Consequently, it is sensitive to underlying abundance gradients, which are present and not well constrained for most extragalactic targets (McConnell et al. 2015). SPS also requires calibration against resolved stellar populations, which are not available for extreme metallicity values.



\subsection{Observational Trends}

The underlying question motivating unresolved studies is whether the IMF varies globally as a function of galaxy type and changes across cosmic time. Due to the challenging nature of the extragalactic observations, studies carried out by various groups using a variety of techniques have historically yielded conflicting, inconsistent and evolving answers to the question of IMF universality. 

In recent years, significant attention has been devoted to studies of ancient early type (elliptical) galaxies (ETGs). These galaxies formed most of their stars in a burst in the early universe with little subsequent star formation. Consequently, their present-day stellar populations are old ($\sim10$ Gyr) and contain only relatively low-mass stars ($M_* \lesssim 1 M_\odot$); they are ``red and dead". Their stellar densities and ages are similar to that of globular clusters, but they have higher average metallicities ($Z \gtrsim 0.5 Z_\odot$). It seems plausible that these extreme conditions hosted significantly different early star formation environments than spiral galaxies like the Milky Way.

Over the last five years a variety of approaches have targeted ETGs and found apparent changes in the IMF for those with the highest stellar velocity dispersions. Spectral feature SPS methods (van Dokkum \& Conroy 2010, van Dokkum \& Conroy 2012, Spiniello et al. 2014), SPS combined with dynamical observations (Callipari et al. 2012, Dutton et al. 2012, McDermid et al. 2014), and SPS combined with gravitational lensing (Treu et al. et al. 2010, Spiniello et al. 2012, Shu et al. 2015) have all concluded that the relative number of low-mass stars increases with increasing velocity dispersion. In concert, these results provide tantalizing evidence for systematic change in the IMF as a function of environment. Confidence is enhanced by the agreement between independent groups using slightly different methods (Conroy et al. 2013). 

Nonetheless, there are outstanding issues. There is substantial scatter in the inferred IMF slopes with some massive ETGs appearing consistent with a Milky Way IMF. It's unclear whether this is due to poorly characterized   uncertainties or arises from actual variation in some additional unidentified physical parameter. There is some evidence that metallicity contributes to the apparent variation (Martin-Navarro et al. 2015, McDermid et al. 2014), but the extent of the dependence is debated.  Upon closer inspection, the methods frequently disagree on the IMF slope of individual galaxies (e.g., Smith \& Lucey 2013, Smith et al. 2015). This suggests that the IMF of individual objects remains very uncertain.

The confluence of uncertainties allows leeway for alternative explanations for the apparent IMF trends. Maccarone (2014) suggests that variations in the Wing-Ford spectral feature used by Conroy \& van Dokkum can be explained by extrinsic S stars (a class of star with enhanced s-process element abundances). These types of stars are neglected by SPS models but could be enhanced in the cores of giant ellipticals. However, it is noted that they might also enhance the calcium triplet line, which is observed to be weak.  If ETGs have bottom-heavy IMFs,  Peacock et al. (2014) note that they should also have smaller populations of neutron stars and black holes, which affect the mass-to-light ratio. They use K-band light to study the low-mass X-ray binary populations of a number of ETGs and find that it is uniform across the sample. This, too, undermines the argument for IMF variation.
Many observations of integrated light have poor resolution within a galaxy and thus average over large areas. Consequently, variation as a function of age or metallicity gradients within the galaxy may masquerade as a varying IMF (McConnell et al.~2015). If gradients in spectroscopy features do correlate with IMF variations, then a bottom-heavy IMF may be limited to a relatively small volume within the galaxy (Mart\'in-Navarro et al. 2015). 
At minimum, this emphasizes that the lingering uncertainties are formidable.

Consequently, there is significant room for variation. Given the age of the stellar populations in ETGs, the integrated light probes a relatively narrow range of stellar masses (e.g., $0.1 M_{\odot} \leq M < 1 M_{\odot}$, Conroy \& van Dokkum 2012). As a result, current evidence for slope variation cannot distinguish between a changing characteristic mass, a changing slope or some combination of the two. This presents a significant challenge for connecting observations with the prior physical conditions of star formation.  Making this connection is essential for building a firm theoretical framework for the IMF and understanding the basis of IMF features.


\section{Progress and Challenges}\label{progress}

In summary, {\it the IMF is surprisingly invariant, but it may vary under extreme conditions.} The apparent robustness of the IMF form both within the Milky Way and across a broad range of extragalactic environments is surprising given variations  in metallicity, gas density, stellar density, and radiative background over several orders of magnitude. There are a few intriguing locations where the IMF may differ. These include the inner parsec of the Galactic center, giant elliptical galaxies (although possibly only their centers), and ultra faint dwarfs. However, it is worth noting that all these environments combined comprise a small fraction of the universe, thus reinforcing that the IMF in the majority of environments and across cosmic time is remarkably uniform.

One major piece of progress is an emerging consensus that the fraction of low-mass stars increases with increasing stellar velocity dispersion. While disagreement continues around the IMF of specific galaxies, this trend is found by several independent methods, including stellar population synthesis, gravitational lensing and dynamical arguments. 
 
 Despite increasing observational agreement, the physical basis for IMF variation remains unconstrained. The underlying physics responsible for star formation and the origin of the IMF is debated. One particular challenge entails linking present-day stellar observables to the natal conditions of star formation.   Power-law fits to the observations are currently incapable of discriminating between a changing characteristic stellar mass and evolution in the relative production of high- and low-mass stars. However, with continuing improvements in stellar models this may be within reach of future work. 
 
 A second challenge involves reconciling inconsistencies in the observations. Trends with velocity dispersion show a large amount of scatter; some ellipticals with very large dispersions appear consistent with the Milky Way IMF. This may be due to uncertainties in the techniques or to some other unidentified variable.  Globular clusters, which resemble massive elliptical galaxies in terms of their stellar densities, display apparently normal IMFs. Determining why globular clusters do not exhibit variation will shed light on their origin and provide revealing clues about the process of star formation, generally.
 
Finally, a comprehensive, predictive theory for the IMF remains elusive. A successful theory must explain both the apparent universality of the IMF across a broad range of environments  and deviation in select extreme environments.  The theory must take into account gas metallicity, cloud properties, and stellar feedback. Until current models succeed in fully explaining star formation within the Milky Way, progress towards understanding extreme environments will be evasive.

\end{document}